\documentclass{aastex61}

\usepackage{newtxtext}
\usepackage{natbib}
\usepackage[version=3]{mhchem} 
\usepackage{graphicx}
\usepackage{verbatim} 

\newcommand\aastex{AAS\TeX}

\received{00/00/00}
\revised{00/00/00}
\accepted{00/00/00}
\submitjournal{ApJ}

\shorttitle{\aastex\ Nucleosynthetic anomalies}
\shortauthors{Jacquet et al.}

\begin{document}

\title{Fingerprints of the protosolar cloud collapse in the Solar System II: Nucleosynthetic anomalies in meteorites}

\correspondingauthor{Emmanuel Jacquet}
\email{emmanuel.jacquet@mnhn.fr}

\author{Emmanuel Jacquet}
\affil{Mus\'eum national d’Histoire naturelle, UMR 7590, CP52, 57 rue Cuvier, 75005, Paris, France}

\author[0000-0003-0902-7421]{Francesco C. Pignatale}
\affil{Mus\'eum national d’Histoire naturelle, UMR 7590, CP52, 57 rue Cuvier, 75005, Paris, France}
\affil{Institut de Physique du Globe de Paris (IPGP), 1 rue Jussieu, 75005, Paris, France}

\author{Marc Chaussidon}
\affil{Institut de Physique du Globe de Paris (IPGP), 1 rue Jussieu, 75005, Paris, France}

\author{S\'ebastien Charnoz}
\affil{Institut de Physique du Globe de Paris (IPGP), 1 rue Jussieu, 75005, Paris, France}



\begin{abstract}

The isotopic heterogeneity of the Solar System shown by meteorite analyses is more pronounced for its earliest objects, the Calcium-Aluminum-rich Inclusions (CAIs). This suggests that it was inherited from spatial variations in different stardust populations in the protosolar cloud. We model the formation of the solar protoplanetary disk following its collapse and find that the solid-weighted standard deviation of different nucleosynthetic contributions in the disk is reduced by one order of magnitude compared to the protosolar cloud, whose successive isotopic signatures are fossilized by CAIs. The enrichment of carbonaceous chondrites in r-process components, whose proportions are inferred to have diminished near the end of infall, is consistent with their formation at large heliocentric distances, where the early signatures would have been preferentially preserved after outward advection. We also argue that thermal processing had little effect on the (mass-independent) isotopic composition of bulk meteorites for refractory elements.

\end{abstract}




\section{Introduction} 

Since the dawn of the Space Age, meteorite analyses have revealed that the Solar System is isotopically heterogeneous in a way that cannot be accounted for by mere mass-dependent fractionations \citep[e.g.][]{JefferyReynolds1961,Claytonetal1973,Leeetal1976,Niemeyer1988,Birck2004,DauphasSchauble2016}. While some of the isotopic anomalies may be due to mass-independent fractionations in a gaseous reservoir \citep[e.g.][]{Youngetal2008} or nuclear reactions within the Solar System, whether by spallation or radioactive decay \citep[e.g.][]{Leeetal1998,Gounelleetal2006, DavisMcKeegan2014,Sossietal2017,Lugaroetal2018,Jacquet2019}, many bear the 
 stamp of presolar stellar nucleosynthesis. Indeed, deviations from terrestrial standards for elements with numerous isotopes may often be readily interpreted in terms of nucleosynthetic contributions such as the s, r and p processes characteristic of different, if sometimes contentious, types of stars \citep{DauphasSchauble2016,Lugaroetal2018}. This ultimately evidences that the protosolar cloud which collapsed to form the solar protoplanetary disk inherited its condensable elements from previous generations of stars. Some of the original stardust, usually submicron-sized, did survive thermal processing in the disk intact and can be identified under the ion probe from its individually \textit{very} anomalous isotopic properties \citep[e.g.][]{Zinner2014}. Such \textit{presolar grains}, it is true, are nowadays a minor component---less than a thousandth of the mass of even the most primitive chondrites---but those that were destroyed passed on their atoms to their environment so that in an effective sense, the isotopic composition of any Solar System body, whether primitive (chondritic) or differentiated, is an average of diverse populations of presolar grains. This very averaging also explains the much more restricted isotopic variations 
 shown by macroscopic meteorite samples compared to their individual interstellar precursors \citep[e.g.][]{NuthHill2004}. 

\begin{figure}[!htbp]
 \begin{center}
 {\includegraphics[width=0.7\columnwidth]{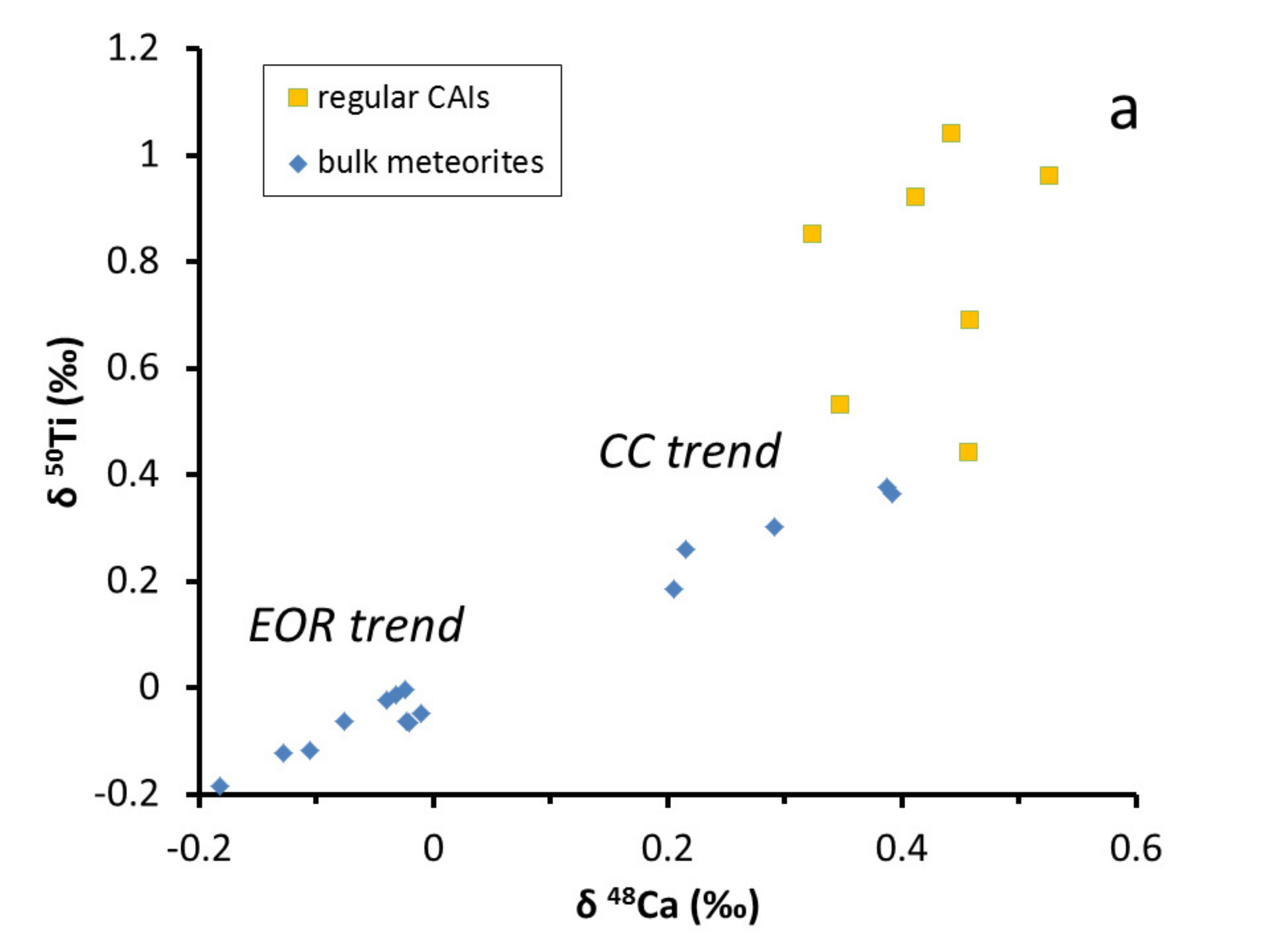}}
 {\includegraphics[width=0.7\columnwidth]{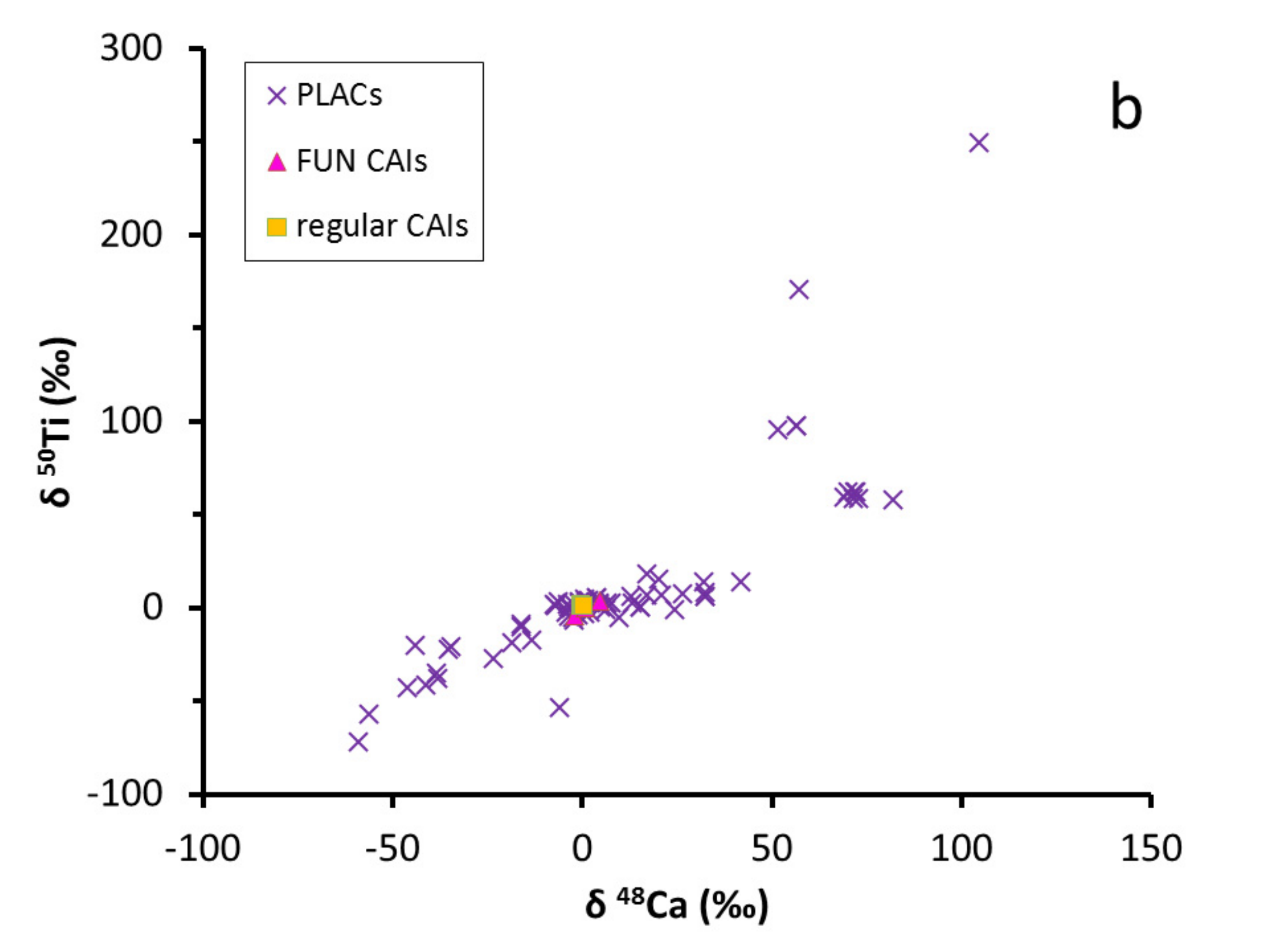}}
 \end{center}
\caption{Isotopic compositions of Ca and Ti for (a) bulk (whole-rock) meteorites (each data point represents the average of a chemical group) and (b) CAIs (with the regular ones also shown on panel a), expressed as relative deviations of  
 $^{48}$Ca/$^{44}$Ca and $^{50}$Ti/$^{47}$Ti from terrestrial standards ($\delta^{48}$Ca and $\delta^{50}$Ti, respectively). Note the difference in scale between the two panels. In panel (a), the ``CC trend'' denotes the carbonaceous chondrites and isotopically affiliated differentiated meteorites; same for ``EOR trend'' and non-carbonaceous chondrites. Modified after \citet{DauphasSchauble2016}, with incorporation of PLAC data from \citet{Koopetal2016} and regular CAI data (from the single chondrite Allende) from \citet{Chenetal2015}.
 } 
\label{Ti vs Ca}
\end{figure}

  Yet, although small, the isotopic deviations shown by bulk (whole-rock) meteorites are far from random (e.g. Fig. \ref{Ti vs Ca}a). A clear hiatus separates the chondrites in two \textit{superclans}, carbonaceous chondrites (CCs) and non-carbonaceous chondrites (EORs, encompassing enstatite, ordinary and Rumuruti chondrites), with the former usually enriched in neutron-rich isotopes \citep{Warren2011b}, in a way correlated with their bulk chemical compositions \citep{Niemeyer1988,Trinquieretal2009}. So there were definite gradients in space and/or time in the proportions of the different presolar contributions accreted in meteorite parent bodies in the disk. Since the space-time ordering of the different chondrite groups, or even the above broad superclans, is anything but
  understood 
  \citep[e.g.][]{Wood2005,Chambers2006, Jacquet2014review,Deschetal2018}, these isotopic systematics represent critical constraints to rationalize.
  
  It is noteworthy that the isotopic anomalies tend to be more pronounced for the earliest Solar System objects. There may be already a discernible temporal trend among bulk meteorites for Ca isotopes \citep{Schilleretal2018} but the greatest anomalies are shown by refractory inclusions, in particular calcium-aluminum-rich inclusions (CAIs), the oldest solids of the solar system \citep{MacPherson2014}. To be sure, CAIs are only one minor chondrite component mixed with others, but inasmuch as they are believed to have originally formed by condensation at high temperatures, they should have been in isotopic equilibrium (some 
  mass-dependent fractionation notwithstanding) with the surrounding gas. Thus, they should represent faithfully the isotopic composition of the region and the epoch where they condensed in the same way a chondrite should represent that of the reservoir where it accreted 
  (as suggested by intragroup isotopic consistency; e.g. \citealt{Pedersenetal2019}). Isotopic anomalies in CAIs are typically one order of magnitude higher than in bulk meteorites, but some minor populations, such as the FUN (``Fractionated and Unknown Nuclear effects'') CAIs and PLACs (PLAty hibonite Crystals) may exhibit anomalies one or two more orders of magnitude higher (see Fig. \ref{Ti vs Ca}b), and those ones are widely believed to represent the very first generation of CAIs \citep[e.g.][]{DauphasSchauble2016,Koopetal2018}. 

  This connection of the isotopic heterogeneity of the Solar System with its earliest epochs suggests a scenario where the protosolar cloud itself was isotopically zoned (i.e. had spatially varying proportions of the different presolar grain populations), and, as it sequentially collapsed to form the solar protoplanetary disk, passed on some of this heterogeneity to it, before turbulent mixing gradually reduced its extent \citep[e.g.][]{HussLewis1995, DauphasSchauble2016,Nanneetal2019}.  In fact, the CAI formation epoch duration appears commensurate with infall timescales \citep{YangCiesla2012,Pignataleetal2018}. Still, many authors in the cosmochemical community argue for differential thermal processing of presolar components in the disk as the cause of isotopic heterogeneities at the bulk meteorite scale \citep[e.g.][]{Niemeyer1988,HussLewis1995,Trinquieretal2009,VanKootenetal2016,Olsenetal2016, Worshametal2019}. Clearly, the scenario of inheritance of large-scale isotopic heterogeneities needs to be explored in a dedicated manner to enable proper evaluation of its merits.

  This is the purpose of this paper. We are engaged in a long-term investigation of the cosmochemical fingerprints of the disk building epoch as a result of the collapse of the protosolar cloud \citep{Pignataleetal2018, Pignataleetal2019}. In \citet{Pignataleetal2019}
  , hereafter ``Paper I'', we already studied the effect of an heterogeneous distribution of the short-lived radionuclide aluminum-26 and its implications on its use as a relative chronometer. The generalization to other ``inherited'' isotopic anomalies, in particular for stable isotopes, was then just at hand. In section \ref{Model}, we present the main features of the model and in particular the transport of isotopic components. Results are presented in section \ref{Results} and compared to the meteoritical record in section \ref{Discussion}. Section \ref{Conclusion} concludes this study.

\section{Model}
\label{Model}

\subsection{Disk building scenario}

  As the numerical implementation of our scenario leans heavily on Paper I, only the briefest of summaries of the disk model will be given here so as to focus on the transport of isotopic anomalies in the most general terms. Unless otherwise noted, the same run parameters as Paper I are used, viz. $T_{cd}=15$~K, $\Omega_{cd}=10^{-14}$ rad.s$^{-1}$, $M_{0,\star}=0.02M_{\odot}$, $T_{\star}=4000$~K, $R_{\star}=3R_{\odot}$, $M_{tot}=1 M_{\odot}$, $\alpha_{active}=10^{-2}$, $\alpha_{dead}=10^{-5}$ (see \citet{Pignataleetal2018} for details).
  
  We consider a protosolar cloud in the form of a singular isothermal sphere \citep{Shu1977} of one solar mass ($M_{tot}=M_\odot$) collapsing from the inside out. At each time during the duration $t_{\rm in}=215$ ka of the collapse period, and following the formulation of \citet{HuesoGuillot2005} the content of a spherical shell infalls (at a rate $\dot{M}_{\rm in}=5\times 10^{-6}\rm M_\odot\cdot a^{-1}$) onto the forming protoplanetary disk inside an heliocentric distance $R_C$ (the centrifugal radius), which increases with time in proportion with the increasing specific angular momentum further away from the cloud center, reaching a maximum of 12 AU under our simulation parameters. As the disk is being built, it also expands viscously beyond the centrifugal radius as a result of angular momentum transport due to turbulence, which is here modelled in an $\alpha$-disk fashion as driven by the magneto-rotational instability, taking into account the existence of a dead zone \citep[e.g.][]{Gammie1996, Zhuetal2010a}.
  
  The code tracks different cosmochemical ``species'' (refractory, main silicate, metal, moderately volatile, ices) characterized by different condensation temperatures (e.g. 1650 K for refractories) above and below which they are wholly gaseous and wholly solid (and mixed with other solids in ``composite'' grains), respectively (see Paper I). In solid form, these species are further divided in ``pristine'' (which never experienced heating above 800 K in the disk), ``condensates'' (which have experienced a gaseous state in the disk) and ``processed'' (all the rest) varieties. At each heliocentric distance, one can thus e.g. distinguish the properties of the condensate fraction from those of the \textit{bulk} (all species and varieties combined). \citet{Pignataleetal2018} could thus see refractory condensates abundantly form early during infall, as matter concentrated on the dense and hot compact young disk, and outward transport of a significant portion thereof following the viscous expansion of the disk (see also \citet{Jacquetetal2011a, YangCiesla2012}).

\subsection{Transport of isotopic anomalies}

  Let us consider a fixed chemical element, or a set of cosmochemically coherent elements (such as refractories as tracked by the code, on which we will focus henceforth). Then the total surface density (including both gaseous and condensed states) of this ``species'' $\Sigma_p$ obeys the mass conservation equation:
\begin{equation}
\label{Sigmap}
\frac{\partial\Sigma_p}{\partial t}+\frac{1}{R}\frac{\partial}{\partial R}\left[R\left(\Sigma_pv_R-D_R\Sigma\frac{\partial}{\partial R}\left(\frac{\Sigma_p}{\Sigma}\right)\right)\right]=\dot{\Sigma}_p
\end{equation}
with $v_R$ the net radial velocity, $D_R$ the turbulent diffusion coefficient, $\Sigma$ the total gas surface density and $\dot{\Sigma}_p$ the infall rate (per unit surface on the midplane) of the species of interest, which is proportional to the total infall rate assuming (to first order) a uniform solar chemistry throughout the cloud. For its gaseous varieties, $v_R$ will coincide with the gas velocity $u_R$, but if solids exist, it must include a drift term owing to finite gas drag \citep{Jacquetetal2012S}. 
In the solid state, the aforementioned drift velocity depends on particle size. While Paper I, as \citet{Pignataleetal2018}, considered a constant fragmentation velocity (on either side of the snow line) when computing the effect of dust coagulation, we choose here to limit the size to 1 mm, which is the likely bottleneck of coagulation suggested by the size distribution of chondrite components \citep{Jacquet2014size}.

  Now, within the species of interest, let us specially consider an isotopic component whose mass fraction therein will be denoted by $x$. By ``isotopic component'', we mean either one particular isotope (e.g. aluminum-26 in Paper I), a combination of isotopes with fixed proportions such as may be produced in a specific nucleosynthetic environment (e.g. s-process molybdenum) or some more complicated mixture thereof. Suffice it to say that the isotopic composition of the species of interest can be expressed as a (linear or homographic) function of the $x$'s of the different isotopic components it contains, so 
   $x$ is a suitable proxy for the evolution of the isotopic composition.
   
  As a concrete example, for Cr isotopes, if we consider $n$ isotopic components of mass fractions $x_1,...,x_n$ (so that $\sum_{i=1}^{n}x_i=1$), where component 1 may refer to ``r-process Cr'', component 2 to ``s-process Cr'' etc., we have:
  \begin{equation}
  \frac{^{54}\mathrm{Cr}}{^{52}\mathrm{Cr}}=\sum_{i=1}^{n}x_i\left[^{54}\mathrm{Cr}\right]_i\diagup \sum_{i=1}^{n}x_i\left[^{52}\mathrm{Cr}\right]_i
  \end{equation}
  with $\left[^{52, 54}\mathrm{Cr}\right]_i$ the concentration (atoms per unit mass) of the stated isotopes in the $i$th component. For small isotopic variations, $^{54}$Cr/$^{52}$Cr can be considered as a linear function of the $x_i$'s, so for practical purposes, one may think of the $x_i$'s as some rescaled isotopic ratios (or linear combinations of isotopic ratios, if we consider other isotopic ratios that lift the degeneracy).

   Assuming that the different isotopic components have the same dynamical behavior\footnote{It would be, in fact, difficult for individual submicron-size presolar grains, if still surviving, to significantly decouple from each other (or from the vapor), and they would be anyway expected to be all rapidly mixed together in millimeter-size aggregates (as in our composite grain approximation; \citet{Pignataleetal2018}), even though \citet{Hubbard2016ferro} entertain fractionation of tungsten isotopes through differential ferromagnetic interactions of their carriers during coagulation. The possibility that isotopic components may be fractionated by evaporation--which is not allowed by our fixed condensation temperature assumption--is discussed in Section \ref{Thermal processing}.}, the surface density of the isotopic component of interest $\Sigma_p x$ obeys the same continuity equation as $\Sigma_p$, such that $x$ evolves as:
\begin{equation}
\label{x}
\frac{\partial x}{\partial t} +\left(v_R-D_R\frac{\partial}{\partial R}\mathrm{ln}\left(\frac{D_R\Sigma_p^2}{\Sigma}\right)\right)\frac{\partial x}{\partial R}-D_R\frac{\partial^2 x}{\partial R^2}=\frac{\dot{\Sigma}_p}{\Sigma_p}\left(x_{\rm in}-x\right)
\end{equation}
where $x_{\rm in}$ is the mass fraction of the isotopic component in the infalling matter. We assume here that each spherical shell of the original protosolar cloud is homogeneous upon arrival on the disk (this will be returned to in section \ref{CAI}) so that on the disk, $x_{\rm in}$ is only a function of time (that is, of original distance from center of the cloud of the infalling matter
). As in Paper I, we will call it the ``injection function''. 

\begin{figure}[!htbp]
 \begin{center}
 {\includegraphics[width=\columnwidth]{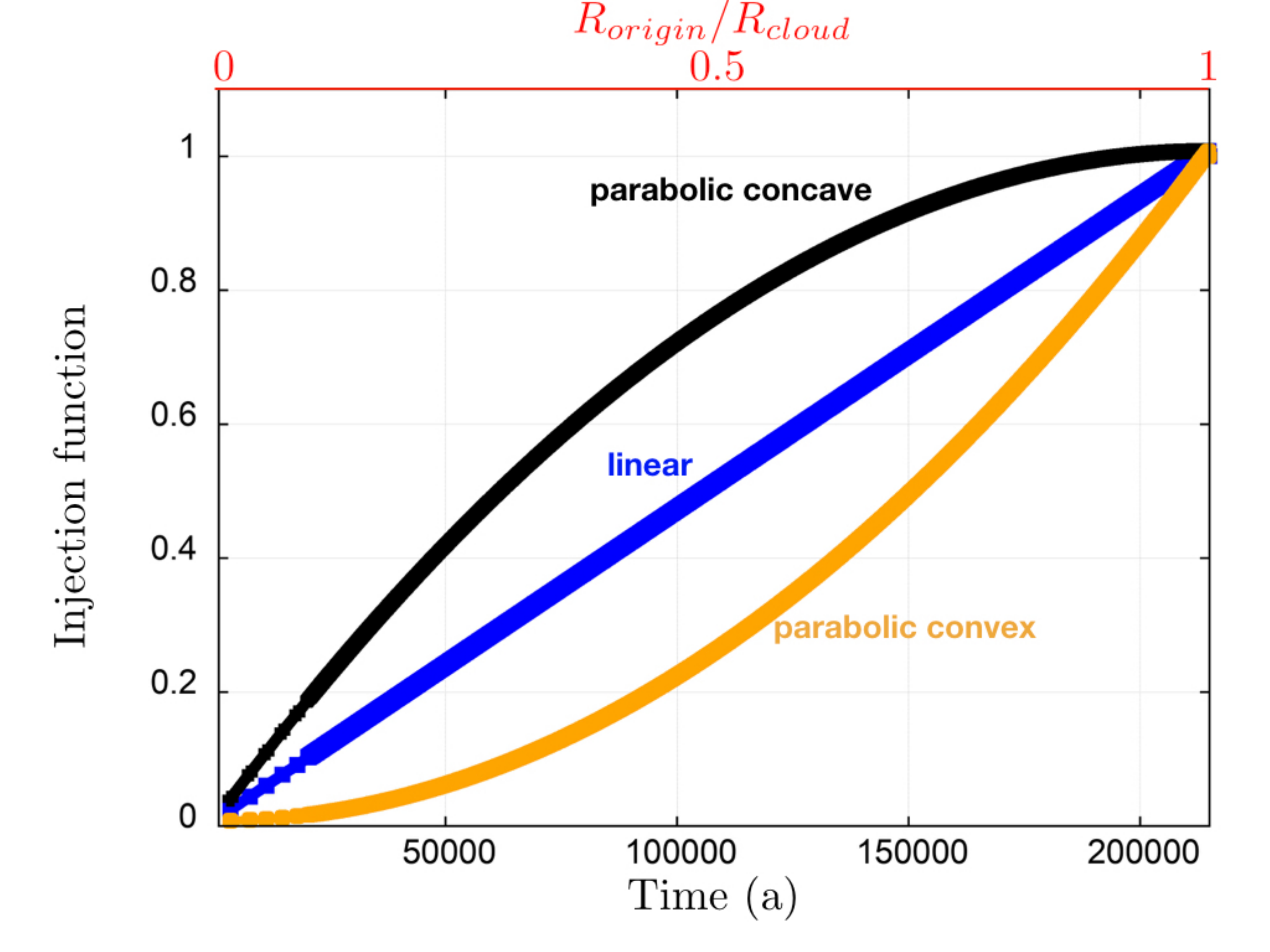}}
 \end{center}
\caption{Plot of the three injection functions (i.e. the normalized isotopic contribution in the infalling material) as a function of time of infall (or equivalently, as shown in the upper x-axis, of original position of the infalling matter in the pre-collapse protosolar cloud).
 } 
\label{xin}
\end{figure}

  Since, for stable isotopes, most relevant nucleosynthetic sources (except for the most recent inputs) may be expected to have spread their products over spatial scales larger than the protosolar cloud, we further specialize to the case where the zoning is monotonic, i.e. $x_{\rm in}$ is a monotonic function of time. It is remembered that Paper I substantiated a monotonic increase for the case of 
  aluminum-26 (we shall nonetheless also return to this assumption in the discussion). Since the solution of equation (\ref{x}) is a linear function of the input $x_{\rm in}$, without loss of generality, we will take $x_{\rm in}$ to span the entire interval [0;1]. We will call the resulting $x$ the ``normalized isotopic contribution'' of the component of interest\footnote{This may e.g. correspond to $(x-\mathrm{min}(x_{\rm in}))/(\mathrm{max}(x_{\rm in})-\mathrm{min}(x_{\rm in}))$. In a simple two-component scenario, if we take the two extreme compositions reached in the cloud as endmembers, it would simply correspond to the contribution of the maximum.}. We can further restrict to increasing injection functions, even if that means a change from $x_{\rm in}$ to its complement $1-x_{\rm in}$. As we have no constraint on the exact shape of the injection functions, we select the simplest possible expressions in our implementations, as follows:
  \begin{itemize}
  \item Linear: $x_{\rm in}=t/t_{\rm in}$
  \item Parabolic convex: $x_{\rm in}=(t/t_{\rm in})^2$
  \item Parabolic concave: $x_{\rm in}=1- (1- t/t_{\rm in})^2$
  \end{itemize}
  These are plotted in Fig. \ref{xin}. The parabolic convex function has an increasing derivative, hence a greater variation near the end of infall, while the reverse holds for the parabolic concave one, the linear function lying in between. This will allow us a glimpse on the effect of the zoning curvature in the cloud.
  
 \section{Results}
 \label{Results}

\begin{figure}[!htbp]
 \begin{center}
 {\includegraphics[width=\columnwidth]{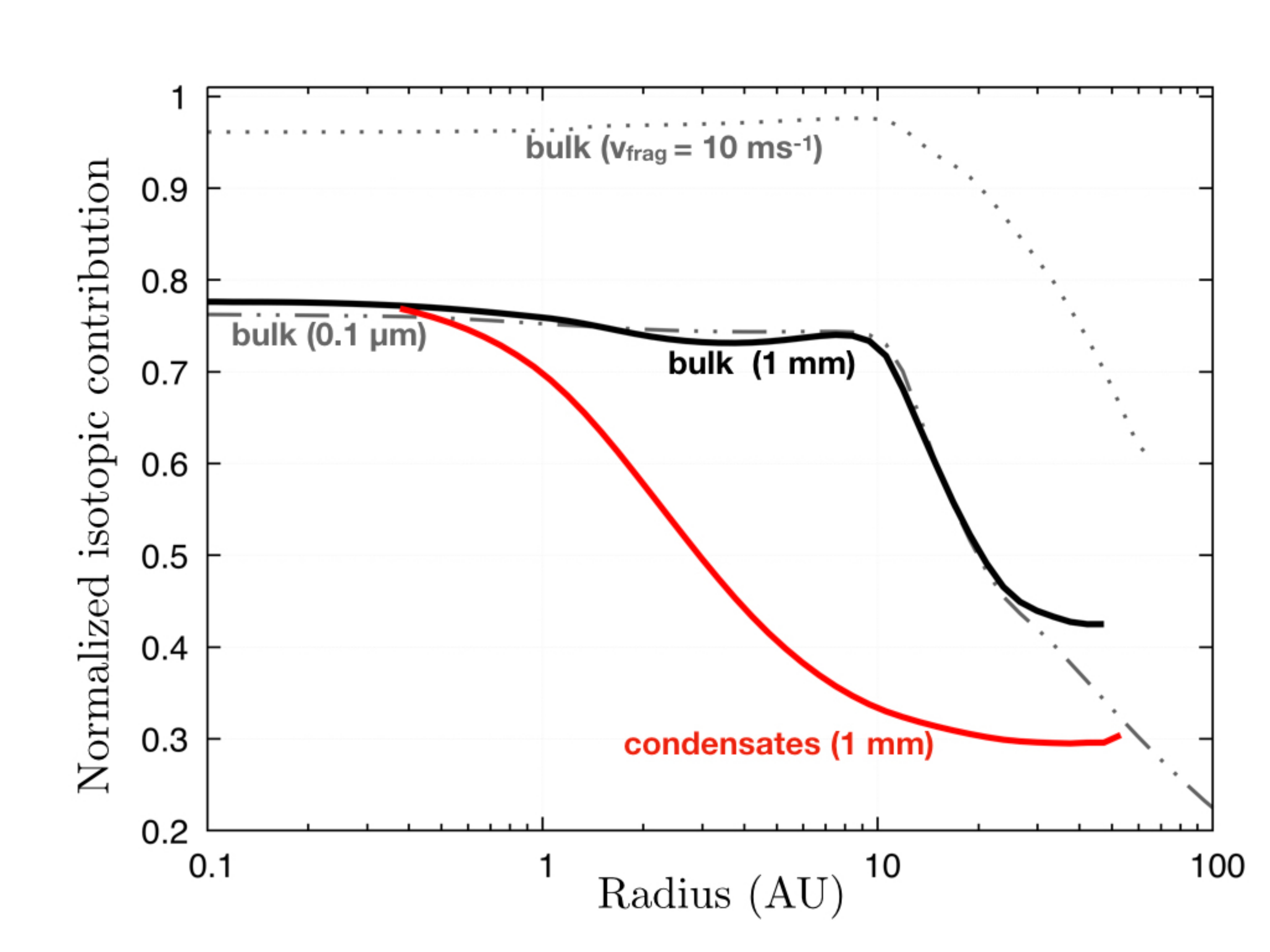}}
 \end{center}
\caption{Plot of normalized isotopic contribution in the bulk (solid thick black line) nebular matter and in the (refractory) condensates (solid red thick black line) for the linear injection function (with the default 1 mm size for solid particles) as a function of heliocentric distance, at the end of infall. Overplotted are also the curves for the bulk in the case of a constant critical fragmentation velocity (dotted line) and a smaller fixed size (0.1 micron; dot-dashed line).
 } 
\label{x vs R}
\end{figure}

In the presentation of the results, we will consider the case with the linear injection function as our fiducial model. Fig. \ref{x vs R} shows the profile of the normalized contribution at the end of infall
. We see that, on the whole, $x$ decreases with heliocentric distance, indicating that the earliest isotopic signatures dominate in the outer disk, despite their having been originally injected in the inner disk (as it had lower specific angular momentum). This is because the viscous expansion of the disk efficiently advected early material outward (see also \citet{Nanneetal2019}) and the non-transported material has been rapidly accreted by the protoSun. There is nonetheless a slight increase discernible just inside the final centrifugal radius (11 AU) where most of the latest (high-$x$) material has been injected, but this washes out during the subsequent ``closed-system'' evolution of the star+disk system. So the effect of viscous expansion is to reverse the gradient of the parental cloud in the disk. The same trend is visible for the refractory condensate composition alone. Their $x$ is systematically lower than that of the bulk (by $\sim 0.2$ with the difference vanishing at the current 
 refractory condensation front), owing to more efficient CAI production during infall than in subsequent times (\citet{YangCiesla2012, Pignataleetal2018}; Paper I).

  Fig. \ref{x vs R} also includes curves for runs performed with (i) the original fragmentation/growth model of \citet{Pignataleetal2018} and (ii) a size limited to 0.1 micron as an assessment of the influence of particle size. The first growth model (i) gives rise to an inner disk much closer to the final infalling composition (see also Paper I). This is because the size of particles may then exceed a few meters and entail rapid migration of the solids from the centrifugal radius inward down to the refractory condensation front. Thus, early signatures are rapidly lost and replaced by later ones for this growth model
  . At the other extreme, a 0.1 micron size makes little difference from our default 1 mm run. This is because millimeter-size solids are already sufficiently small to be tightly coupled to the gas \citep{Jacquetetal2012S}, which is \textit{a fortiori} the case for those. This implies that the volatility of the considered element has little effect on the isotopic zoning of the bulk disk, since whether a nucleosynthetic component is condensed or not, it will behave like the gas on large scales. Thus, in principle, our results could apply to non-refractory elements (but see Section \ref{CAI} for limits on this generalization).

  \begin{figure}[!htbp]
 \begin{center}
 {\includegraphics[width=\columnwidth]{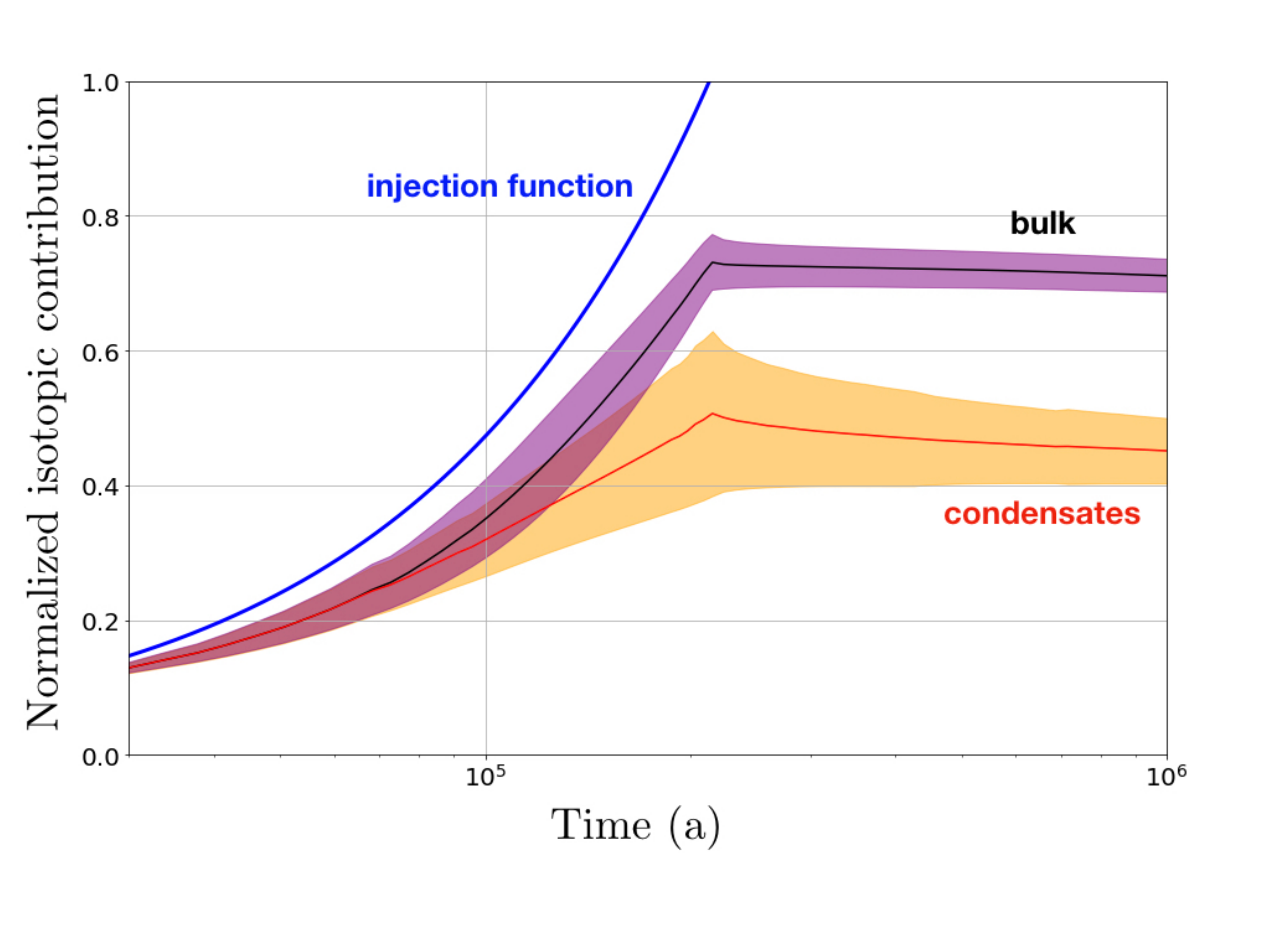}}
 \end{center}
\caption{Plot of the average normalized isotopic contribution in the bulk and (refractory) condensates as a function of time, along with standard deviation represented by shading, for the linear injection function (also plotted; beware the semi-logarithmic scale). Averages and standard deviations are weighted according to the solids in the disk.  The average roughly follows the injection function, but with some lag (greater for the condensates), and evolves only slightly after cessation of infall, with a steadily decreasing standard deviation.} 
\label{avg stdev}
\end{figure}

  \begin{table*}[!htbp] 
  \begin{center}
\begin{tabular}{|c|c|c|c|c|}

\cline{1-4} 

Injection function  & Protosolar cloud & Disk (end of infall) & Disk (t=1 Ma) \\
\hline 
Linear & 0.5 $\pm$ 0.289 & 0.731 $\pm$ 0.041 & 0.711 $\pm$ 0.024 & bulk\\
 &  & \textit{0.507 $\pm$ 0.122} & \textit{0.451 $\pm$ 0.048} & \textit{condensates}\\
\hline
Parabolic convex & 0.333 $\pm$ 0.298  & 0.621 $\pm$ 0.053 & 0.609 $\pm$ 0.029 & bulk\\
 &   & \textit{0.388 $\pm$ 0.149} & \textit{0.406 $\pm$ 0.082} & \textit{condensates}\\
\hline
Parabolic concave & 0.667 $\pm$ 0.298  & 0.881 $\pm$ 0.033 & 0.863 $\pm$ 0.020 & bulk\\
 &  & \textit{0.689 $\pm$ 0.107} & \textit{0.637 $\pm$ 0.045} & \textit{condensates}\\
\hline
\end{tabular} 
\caption{Solid-weighted average ($\pm$ standard deviation) normalized isotopic contribution in the disk (at the end of infall and the end of the simulation) compared to the original cloud. For each injection function, the first row refers to the bulk matter and the second (in italics) to the sole (disk) condensates. Note that the standard deviation for the condensates only represents the variation of the mean value accross the different heliocentric distances, and not the dispersion of individual condensates in a chondrite, which our Eulerian code does not access.}
\label{stdev}
\end{center}
\end{table*}

    While the simulations show a significant spread of the normalized isotopic contribution throughout the disk, the largest heliocentric distances actually contain little mass. A better proxy of the meteoritical record is to consider average and standard deviations \textit{weighted by the mass of solids}, as shown in Fig. \ref{avg stdev}. Representative values for the three injection functions are tabulated in Table \ref{stdev}. It is seen that the standard deviation is always much smaller than that of the original protosolar cloud. Indeed, early signatures are rapidly transported and end up to a large fraction in the protostar, so the average composition of the disk will be biased toward the latest contributions of infall. This is especially the case for injection functions which vary little near the end, as may be seen when comparing the ``parabolic concave'' one (which has the lowest standard deviations) with the others. Still, as a rule of thumb, the reduction factor of the standard deviation between the cloud and the disk is always about one order of magnitude for the different injection function shapes. This may be connected to the fact that at cessation of infall, the disk essentially represents the outermost few tenths of the cloud in mass ($\sim$0.3 here, see Paper I), with some turbulent homogenization. The disk turnover timescale itself $t_{\rm vis}=R/|u_R|$ 
     evaluated at the disk edge $R=R_D$, represents a similar fraction of the total infall time $t_{\rm in}=M_\odot/\dot{M}_{\rm in}$ since:
    \begin{equation}
  \frac{t_{\rm vis}(R_D)}{t_{\rm in}}=\frac{\dot{M}_{\rm in}}{\dot{M}}\frac{2\pi R_D^2\Sigma(R_D)}{M_\odot}\sim \frac{M_D}{M_\odot}
    \end{equation}
    with the mass accretion rate through the disk $\dot{M}=-2\pi R\Sigma u_R$ and where the approximation uses $\dot{M}\sim\dot{M}_{\rm in}$ and $M_D\sim 2\pi R_D^2\Sigma(R_D)$ for the disk mass. 
    
    The estimate $M_D/M_{\rm tot}\sim 10^{-1}$ (the reciprocal of the standard deviation reduction factor) should be fairly robust. If we consider the disk surface density to be a power-law $\Sigma\propto R^{-p}$ up to $R_D$, equating the angular momentum carried by the disk to that in the original (singular isothermal sphere) cloud yields a disk-to-total mass ratio:
    \begin{equation}
    \frac{M_D}{M_{\rm tot}}=\frac{5/2-p}{3(2-p)}\left(\frac{R_C}{R_D}\right)^{1/2}
    \end{equation}
    which should not be much lower than $10^{-1}$ unless the disk expanded \textit{very much} beyond the centrifugal radius by then. Independently of the idealizations of our hydrodynamical model, the ``minimum mass solar nebula model'' \citep{Hayashi1981} anyway provides a strong lower limit for $M_D$ around $0.02 M_\odot$ for the Solar System.

    After cessation of the infall, the standard deviation decreases and the disk average evolves toward slightly lower (earlier) values as inner disk material, which tends to carry later signatures (high $x$), is progressively lost to the Sun. So there is a change in the direction of variation at cessation of infall, since this average was previously following, if with some lag, the increase of the injection function. 

\begin{figure}[!htbp]
 \begin{center}
 {\includegraphics[width=\columnwidth]{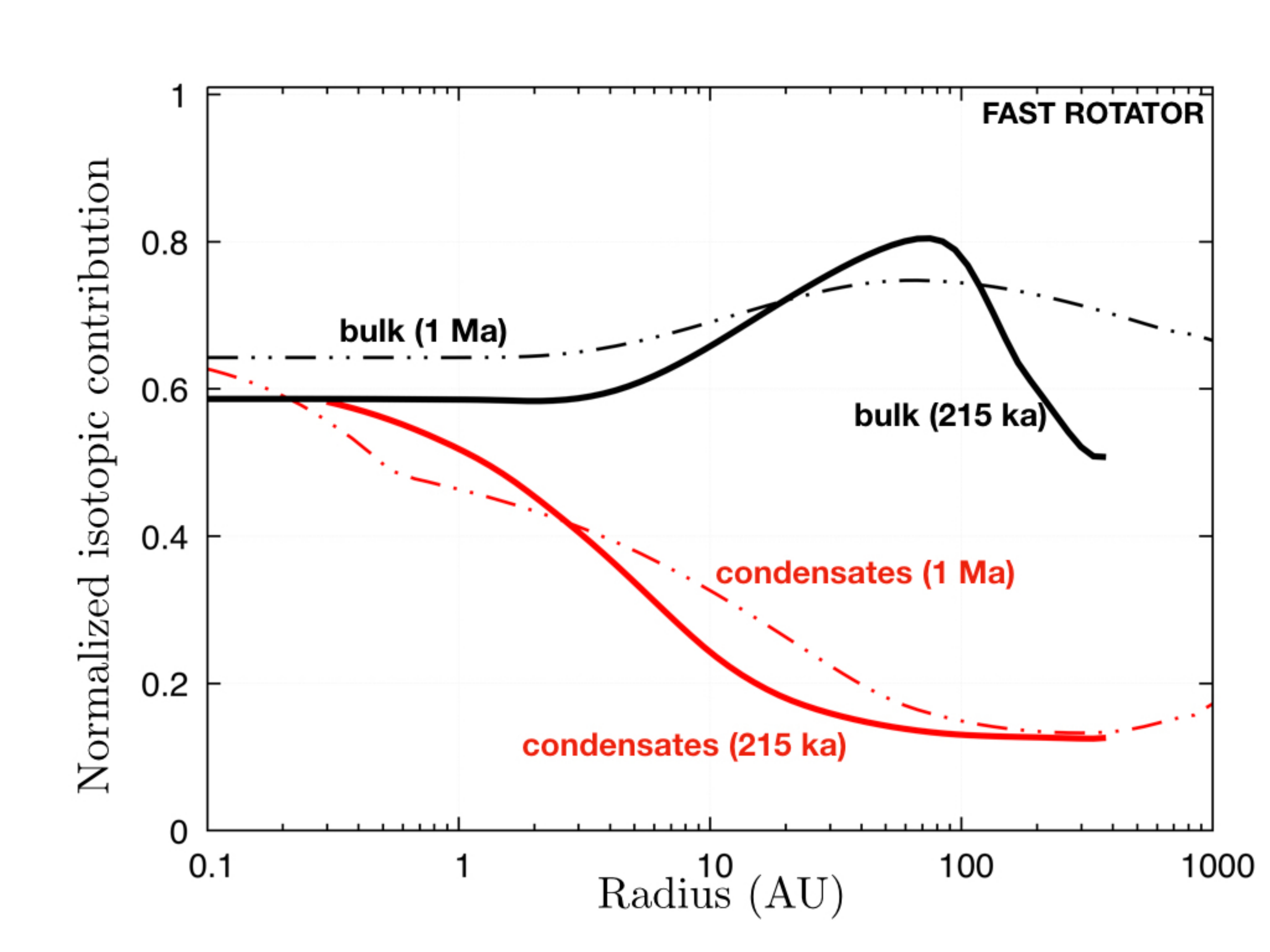}}
 \end{center}
\caption{Plot of normalized isotopic contribution in the bulk (black) nebular matter and in the (refractory) condensates (red), at cessation of infall (continuous) and the end of the simulation (t=1 Ma), for the linear injection function and faster rotation of the protosolar cloud.
 } 
\label{fast x vs R}
\end{figure}

  We also ran simulations with a three times faster rotation of the protosolar cloud. This implies a larger maximum centrifugal radius of 98 AU. As may be seen in Fig. \ref{fast x vs R}, owing to longer transport timescales $t_{\rm vis}(R_C)$, the normalized isotopic contribution rise near this radius for the bulk contribution is more pronounced at the end of infall and does not wash out during simulation time. Hence, the inner disk gradient (and time evolution) is opposite to the former case, although that of the condensates 
  still exhibits a monotonic decrease. On the whole, the isotopic signatures in the disk are closer (by $\sim 0.1$; see Table \ref{stdev_fast}) to the average of the protosolar cloud since the earliest ones have been less concentrated in the dangerous vicinity of the accreting protoSun.

 \begin{table*}[!htbp]
\begin{center}
\begin{tabular}{|c|c|c|c|c|}
\cline{1-4} 
 Injection function & Protosolar cloud & Disk (end of infall) & Disk (t=1 Ma) \\
\hline
Linear & 0.5 $\pm$ 0.289 & 0.612 $\pm$ 0.047 & 0.659 $\pm$ 0.024 & bulk\\
& & \textit{0.384 $\pm$ 0.108} & \textit{0.399 $\pm$ 0.065} & \textit{condensates}\\
\hline
Parabolic convex & 0.333 $\pm$ 0.298 & 0.466 $\pm$ 0.064 & 0.533 $\pm$ 0.036 & bulk\\
& & \textit{0.237 $\pm$ 0.094} & \textit{0.256 $\pm$ 0.063} & \textit{condensates}\\
\hline
Parabolic concave & 0.667 $\pm$ 0.298 & 0.802 $\pm$ 0.033 & 0.836 $\pm$ 0.015 & bulk\\
& & \textit{0.538 $\pm$ 0.118} & \textit{0.561 $\pm$ 0.079} & \textit{condensates}\\
\hline
\end{tabular} 
\caption{Same as Table \ref{stdev} but for a faster protosolar rotation $\Omega_{cd}=3\times 10^{-14}$ rad.s$^{-1}$.}
\label{stdev_fast}
\end{center}
\end{table*}

\section{Discussion}
\label{Discussion}

\subsection{Refractory inclusions as probes of the infall stage}
\label{CAI}

   In our model, the isotopic variability of refractory inclusions should reflect the temporal evolution of the isotopic composition of the refractory condensation front during infall, and somewhat beyond (although the latest CAIs would not be preferentially preserved; \citet{YangCiesla2012}). On average, CAIs should retain isotopic signatures relatively more ``archaic'' (in terms of the evolution of the infalling matter) than their hosts. Since they tend to be enriched in r-process contributions (ascribed to neutron-rich environments such as supernovae; \citet{Lugaroetal2018}) compared to bulk meteorites \citep[see e.g.][ Fig. \ref{Ti vs Ca}a]{DauphasSchauble2016}, this indicates that the injection function of r-process contributions decreased during infall. This has also been suggested by \citet{Nanneetal2019,Burkhardtetal2019}. So we may think of the $x$ (increasing with time) plotted in the paper to represent the complement of the r-process component, if these runs are to be likened to the forming solar system.

  FUN CAIs would then represent CAIs generally earlier than regular CAIs, 
 which would agree roughly with the order of magnitude difference in standard deviation expected between the cloud and the post-infall disk (a few \textperthousand~  
  vs. a few parts per 10,000 (\textpertenthousand); \citet{DauphasSchauble2016})
 . PLACs however show even more variability, with e.g. a standard deviation for $\delta^{50}$Ti of 47 \textperthousand 
  for the \citet{DauphasSchauble2016} compilation vs. 0.2 \textperthousand~ 
   for bulk meteorite groups (from the compilation of \citet{DauphasSchauble2016}; adding PLAC analyses of \citet{Koopetal2016} and bulk data of \citet{Schilleretal2018}). More troublesome is that while the average of +$12\pm 6$ \textperthousand 
    lies on the same side of the meteorite bulks as regular CAIs, they scatter almost symmetrically on both sides thereof (Fig. \ref{Ti vs Ca}b; in fact some such scatter is already apparent for FUNs). 
 It would be conceivable that the injection function varied nonmonotonically (e.g. because of turbulence interchanging parcels from a large-scale gradient; see e.g. Appendix \ref{Paper I injection}), but there is no reason it should have been by an amplitude so much greater than in the final stage, as the molecular cloud would have been overturned over scales much greater than our primeval dense core. Besides, it would seem quite coincidental that the 
 disk, largely sampling the \textit{outer} portion of the cloud, ended up so close to the average of the \textit{inner} cloud (within one tenth of its dispersion) despite such variations. 
 
   So the variability of PLACs may instead reflect spatial and/or short-timescale variations of the injection functions not captured by our 1D model, where $x_{\rm in}$ is a smooth function solely of time because each infalling (spherical) shell is assumed to be homogeneous. Latitudinal variations within each shell of the protosolar cloud, leading to variations of  $x_{\rm in}$ with heliocentric distance, would likely not change the picture much as CAIs would condense at essentially one heliocentric distance, the (axisymmetric) refractory condensation front, but heterogeneities in the azimuthal directions (or more realistic infall scenarios) would produce transient isotopic variations along it. Indeed, The timescale $t_{\rm az}$ for azimuthal mixing by turbulent diffusion and Keplerian shear may be estimated by setting:
   \begin{equation}
  \frac{3}{2}\Omega \frac{\sqrt{D_Rt_{\rm az}}}{R}t_{\rm az}=2\pi
  \end{equation}
 with $\Omega$ the Keplerian angular velocity, hence:
  \begin{eqnarray}
  t_{\rm az}=\Omega^{-1}\left(\frac{4\pi}{3}\frac{\Omega R}{c_s}\right)^{2/3}\delta^{-1/3}
=24\:\mathrm{a} \left(\frac{R}{1\:\rm AU}\right)^{7/6}\left(\frac{1500\:\rm K}{T}\right)^{1/3}\left(\frac{10^{-3}}{\delta}\right)^{1/3}
  \end{eqnarray}
  where $D_R\equiv\delta c_s^2/\Omega$, with $c_s$ the sound speed and $T$ the temperature. This is comparable to the timescales of homogenization of an injected color field in a marginally gravitationally unstable disk simulated by \citet{Boss2007}. Thus, we might imagine that discrete isotopically distinctive cloud parcels landed on the midplane, and that the earliest condensates (or evaporation residues, which would then sample the variability of individual interstellar dust aggregates) such as refractory hibonite fossilized their isotopic signatures before homogenization. As the lower star/disk mass ratio {at that time} would indeed make the disk proner to gravitational instabilities, one could speculate that those discrete parcels induced local collapses of disk regions, hereby retaining longer their isotopic individuality, whose resulting heating may have attained CAI-forming temperatures \citep[e.g.][]{Nayakshinetal2011}. This means that a smaller-scale spatial heterogeneity may be entangled with the secular evolution of the (shell-wide) average of the infalling matter for individual analyses. This does not hinder the identification of the secular trends, with sufficient statistics, as above when inferring the r-process diminution near the end of infall from CAI-bulk comparison, or as in Paper I when ascribing the near-systematic depletion in $^{26}$Al of PLAC and FUN CAIs, with no ``complementary'' super-canonical value, to the lack of $^{26}$Al (above background) near the core of the protosolar cloud.
 
   Regardless, the PLACs (or other CAI populations) would still provide a good proxy for the isotopic heterogeneity of elements in the disk, for we may expect that the reduction factor in standard deviation (due to immediate and long-term 
    mixing in the disk) between them and bulk meteorites was of the same order of magnitude for different isotopic ratios. We may verify a factor of $\sim$200 applies for both $\delta^{50}$Ti (see above) and $\delta^{48}$Ca (34 vs. 0.2 \textperthousand~ 
     from the same compilation), but isotopic data for other refractory elements in PLACs would be worthwhile. As we argued above that the model should in principle apply to nonrefractory elements, it is worth noting that the few \textperthousand~ relative diversity in $^{26}$Mg/$^{25}$Mg ratios in PLACs \citep[e.g.][]{Liuetal2012} would translate in tens of ppm diversity in bulk meteorites independently of $^{26}$Mg ingrowth by $^{26}$Al decay. This is similar to the discrepancies observed by \citet{Olsenetal2016} with respect to an homogeneous $^{26}$Al distribution scenario, so $^{26}$Al homogeneity after infall needs not be cast in doubt
    , whatever its injection function may have been (see Paper I). PLACs also exhibit O isotopic diversity with a 3 \textperthousand~ standard deviation in the mass-independent parameter $\Delta^{17}$O \citep{Koopetal2016}. This may likewise trace an heterogeneity in nucleosynthetic contributions in the protosolar cloud (or self-shielding effects therein as in \citet{Yangetal2011}, as our model is indifferent to the origin of the zoning in the protosolar cloud as long as it is passively inherited). While spinel-hibonite spherules (SHIBS), which show smaller nucleosynthetic anomalies, are tightly clustered around the present solar isotopic composition of oxygen \citep{Koopetal2016}, bulk meteorites are systematically $^{16}$O-depleted by about 50 \textperthousand, with 1.7 \textperthousand~  standard deviation \citep{DauphasSchauble2016}. This is usually ascribed to interactions between isotopically distinct silicates and gas (e.g. water), whether the difference is inherited from the protosolar cloud \citep{Krotetal2010}\footnote{The variations in the disk would however rely on strong solid/gas fractionations, despite their difficulty \citep[e.g.][]{HubbardMcLow2018}} or some fractionation process in the gas phase \citep{ThiemensHeidenreich1983,Youngetal2008}, which may have yielded $^{16}$O-poor water (as evidenced from aqueous alteration products; \citealt{Sakamotoetal2007,Vacheretal2017}). Clearly, for so volatile an element as O, our treatment of instantaneous condensation will break down, with comparable proportions existing in the gas and the solids over a large temperature interval, allowing significant drift, while sluggish kinetics and/or fractionations at low temperatures will not allow to consider rocks as representative of the isotopic composition of the reservoir where they formed or accreted.

\subsection{The carbonaceous/non-carbonaceous chondrite dichotomy}

  If, as argued before, r-process contributions tended to decrease near the end of infall, CCs, which are more enriched in those, should have accreted later and/or further from the Sun than the EORs owing to the inversion of the parental cloud gradient in the disk. The latter ``common wisdom'' spatial interpretation is consistent with the general stratification of the asteroid belt, with S-complex asteroids, associated with ordinary chondrites \citep{Nakamuraetal2011}, generally more numerous in its inner parts, whereas the C-complex, with rather CC affinities,  dominates further out \citep[e.g.][]{DeMeoCarry2014}. This is however directly opposite to the \citet{Jacquetetal2012S} conjecture that chondrite components were more tightly coupled to the gas for CCs than for EORs, so as to explain, \textit{inter alia}, the greater abundance of refractory inclusions and matrix-chondrule complementarity in the former. 
  
\begin{figure}[!htbp]
 \begin{center}
 {\includegraphics[width=\columnwidth]{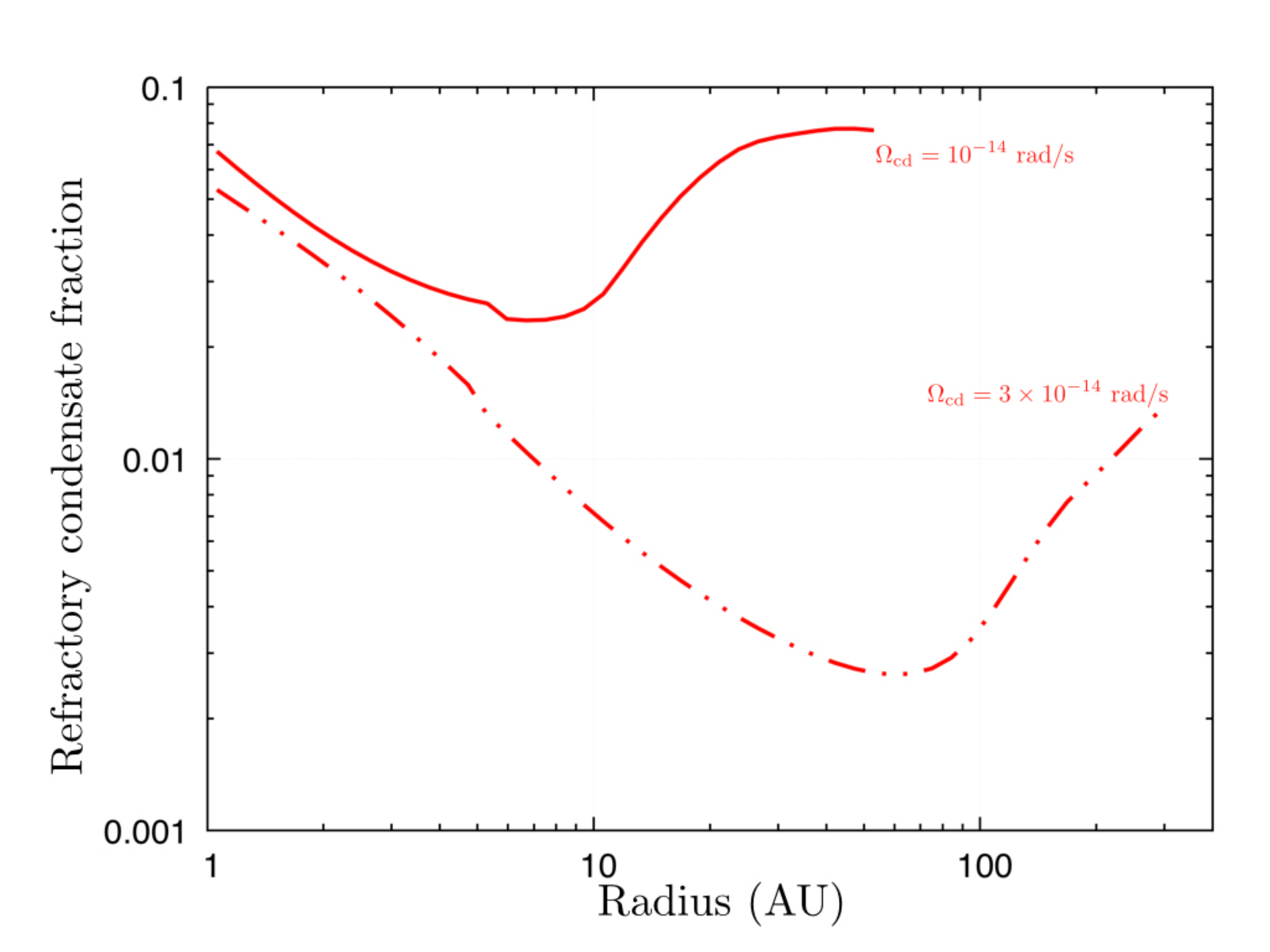}}
 \end{center}
\caption{Fraction of refractory condensates in solid matter (beyond the main silicate condensation front) at cessation of infall, for our default (solid line) and enhanced (dash-dotted line) rotation rate of the protosolar cloud. The former case (corresponding to a compact initial disk) allows efficient outward advection of refractory condensates and an inverted gradient.
 } 
\label{CAI abundance}
\end{figure}

  In principle, one could evade the contradiction by denying the inversion of the parental cloud gradient \citep{Visseretal2009,Nanneetal2019}. This could come about for a greater centrifugal radius, as in our faster rotator runs (which would however greatly diminish the abundance of refractory inclusions; see Fig. \ref{CAI abundance}; \citet{Ciesla2010, YangCiesla2012}), or suitable variations of the $\alpha$ parameter in the outer disk. One would then expect CAIs in non-carbonaceous chondrites to carry more ``archaic'' signatures than in carbonaceous chondrites. However, if anything, CAIs in ordinary chondrites seem to be \textit{less} r-process enriched than in CCs, with an average $\delta^{50}$Ti of $0.6\pm 0.1$ 
   \citep{Ebertetal2018}, as compared to $0.96\pm 0.03$ \textperthousand
   for CV chondrite CAIs \citep{Torranoetal2018}, although data are few\footnote{We do not, however, consider the Na-Al-rich chondrules analyzed by those authors as the result of melting of isotopically ordinary chondrite-like CAIs since, notwithstanding their enrichment in volatile Na, their REE patterns reported by \citet{EbertBischoff2016} resemble more those seen in enstatite chondrite chondrules \citep{Jacquetetal2015EC} than group II or III ones seen in CAIs (save for object Dha1-6, whose Ti isotopes have however not been measured), suggesting a similar origin, probably acquired during chondrule formation rather than inherited from a refractory precursor.} 
    . More measurements of nucleosynthetic anomalies in non-carbonaceous chondrite CAIs would be obviously desirable.
    
    If viscous expansion did bring about the inversion of the parental cloud gradient, it may also account for the enrichment of CAIs in the outer disk (Fig. \ref{CAI abundance}; \citet{YangCiesla2012,Pignataleetal2018}). Possibly, Jupiter may have carved an annular gap in the disk, hence a pressure maximum at its outer boundary, which may have concentrated CAIs as well \citep{Deschetal2018}. Any of the above would alleviate the CAI argument of \citet{Jacquetetal2012S}. Still, evidence for matrix-chondrule complementarity in CCs rather than EORs would remain challenging \citep[e.g.][]{Goldbergetal2015}. A long way remains to understand chemical fractionations of chondrites \citep{Jacquet2014review}.

\begin{figure}[!htbp]
 \begin{center}
 {\includegraphics[width=\columnwidth]{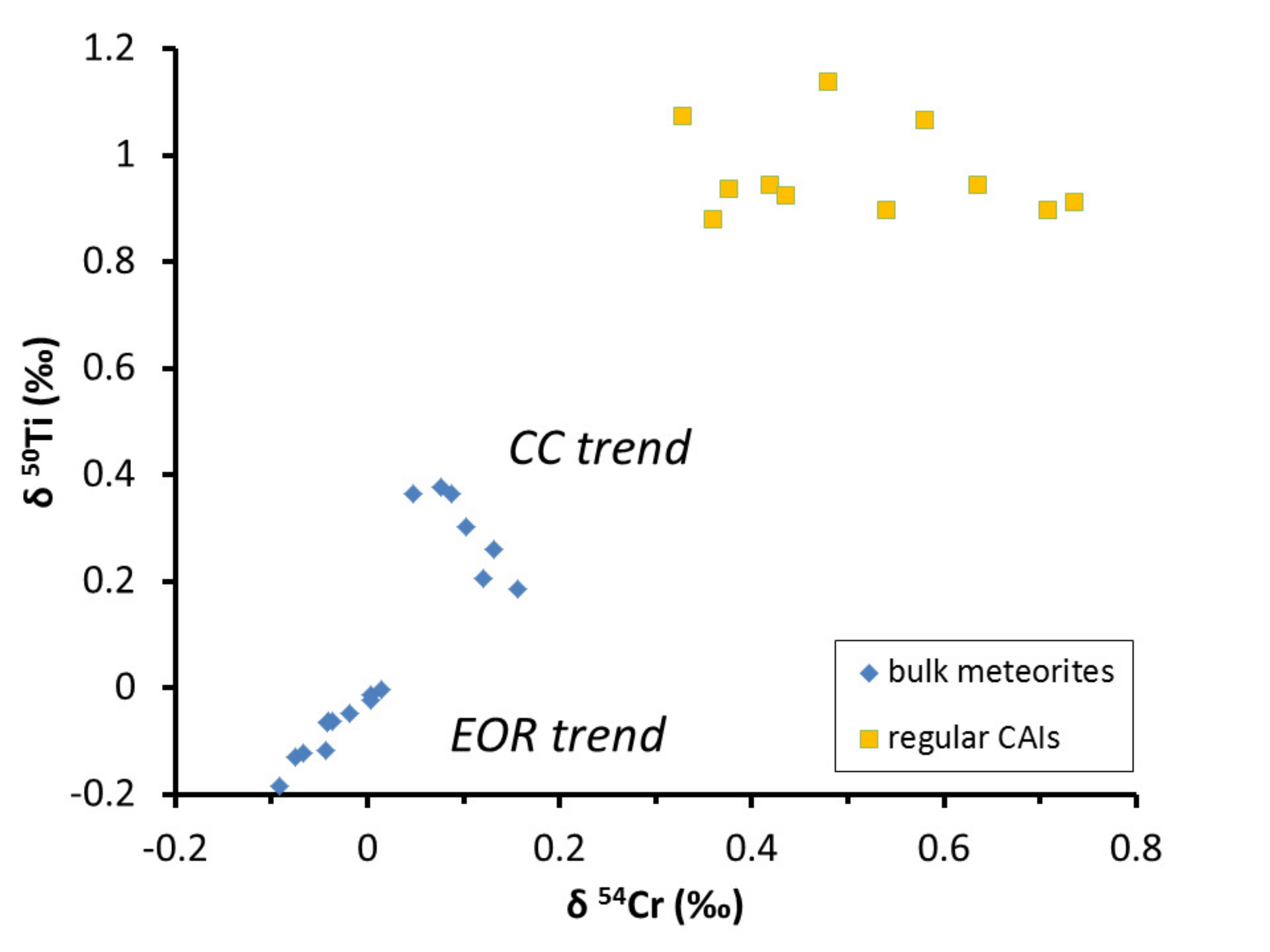}}
 \end{center}
\caption{Isotopic compositions of Cr and Ti for bulk (whole-rock) meteorites and regular CAIs, expressed as relative deviations of  
 $^{54}$Cr/$^{52}$Cr  and $^{50}$Ti/$^{47}$Ti ratios from terrestrial standards ($\delta^{54}$Cr and $\delta^{50}$Ti, respectively). Data from \citet{DauphasSchauble2016} for bulk meteorites and \citet{Torranoetal2018} for regular CAIs (from CV chondrites).
 } 
\label{Ti vs Cr}
\end{figure}

  Our simulations return isotopic compositions which continuously vary with time and heliocentric distance in the solar protoplanetary disk. Certainly, the meteoritical record would not be expected to exhibit an isotopic continuum given the severe depletion of the asteroid main belt where only one out of a thousand or so planetesimals was spared from ejection or incorporation in planets \citep{Jacquetetal2016}. A hiatus between CCs and EORs in isotopic space is however undeniable (Fig. \ref{Ti vs Ca}a; Fig. \ref{Ti vs Cr}) and does indicate an incompleteness of the model. A popular add-on in recent literature is to consider that the aforementioned Jupiter-carved gap pressure maximum filtered out dust drifting inward past it \citep[e.g.][]{Deschetal2018, Nanneetal2019}, sealing off the inner disk from the outer disk. It is uncertain, though, whether the dust flow would be significantly interrupted given the size sensitivity of this concentration process \citep{Haugbolleetal2019}. Another possibility could be that Jupiter (and Saturn?) efficiently accreted planetesimals with intermediate isotopic compositions, removing them from the meteoritical record. Then, the existence of the isotopic dichotomy at the time of accretion of iron meteorite parent bodies \citep{Kruijeretal2017} does not necessarily imply that Jupiter formed within the first Ma of the Solar System, for the intermediate compositions might have been removed later. In this respect, the \citet{Kruijeretal2017} results would merely mean that the observed isotopic spread already existed early on (which, according to our model, would be a necessity as it would but decrease with time). How interesting a sample from a regular satellite of Jupiter would be!
  
\begin{figure}[!htbp]
 \begin{center}
 {\includegraphics[width=\columnwidth]{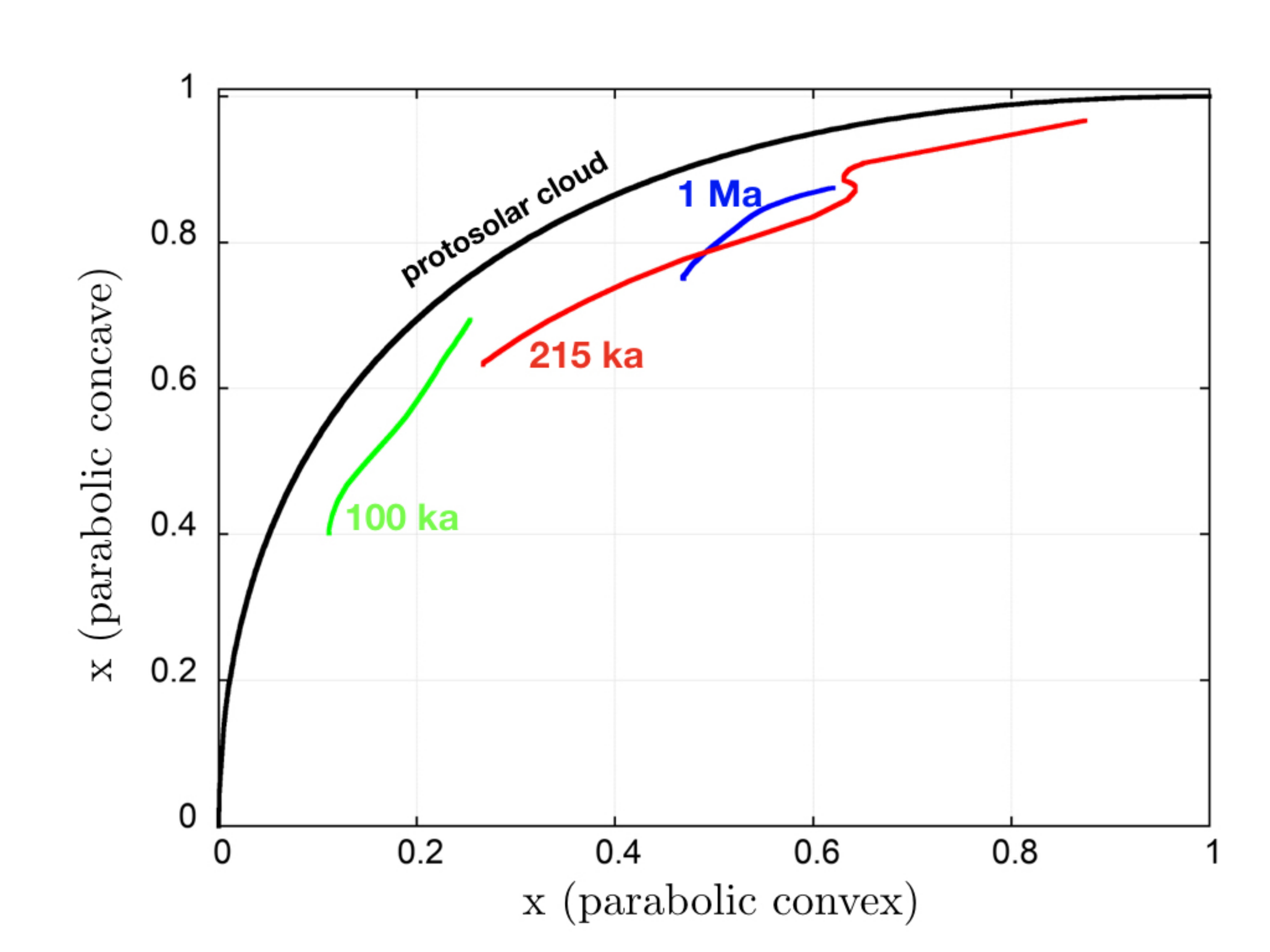}}
 \end{center}
\caption{Biplot of normalized isotopic contributions corresponding to two different injection functions (parabolic convex and parabolic concave) for three different times. Basically, each curve shows how two different isotopic ratios (governed by different nucleosynthetic contributions, as in Fig. \ref{Ti vs Cr}) co-vary accross the various locations of the disk at the stated epoch. It is seen that the functional relationship between the two changes with time and differs from the original one in the protosolar cloud, also plotted.
 } 
\label{concave vs convex}
\end{figure}

  The carbonaceous/non-carbonaceous chondrite dichotomy has another more puzzling aspect. While plots such as Fig. \ref{Ti vs Ca}a simply suggest an incomplete, but well-defined single trend of co-variation between isotopic ratios, other biplots show quite different trends for the two chondrite superclans. Consider for example Fig. \ref{Ti vs Cr}: the EORs show a positive correlation between $\delta^{54}$Cr and $\delta^{50}$Ti while these anticorrelate among CCs
  . Regular CAIs exhibit yet another, horizontal, trend. This is intriguing, as in the simplest case of a monotonic variation of the normalized isotopic contribution in the interstellar medium, one would expect monotonic co-variations in the molecular cloud (as in Fig. \ref{Ti vs Ca}b), even if turbulence interchanges parcels. Interestingly, this kind of divergence is seen for isotopic ratios for the single element molybdenum---arguing against a role of the chemistry---, where CCs seem to be globally enriched in a r-process contribution compared to EORs, but internal variations within each of these superclans seem to mark variable proportions of an s-process contribution \citep{Buddeetal2016Mo}. This admittedly could suggest to decouple the explanations for the two trends (inter- and intra-superclan), as in the redox-dependent thermal processing scenario of \citet{Worshametal2019}. However, under the above hypothesis of a monotonic co-variation of some isotopic ratios in the protosolar cloud, it does not actually follow that the same co-variation should endure in the disk. Indeed, if a second normalized isotopic contribution $X$ can be expressed (at least locally) as a function $f$ of $x$ in the disk, this function must depend on time (i.e. be bivariate: $X=f(t,x)$). Indeed, by manipulating equation (\ref{x}) applied to $X$: 
\begin{eqnarray}
\frac{\partial f}{\partial t} &=& \frac{\dot{\Sigma}_p}{\Sigma_p}\left(F(x_{\rm in})-f(t,x)-(x_{\rm in}-x)\frac{\partial f}{\partial x}\right) + D_R \left(\frac{\partial x}{\partial R}\right)^2\frac{\partial^2 f}{\partial x^2},
\end{eqnarray}
where $F$ represents the original functional relationship in the cloud (i.e. $X_{\rm in}=F(x_{\rm in})$). This amounts to a diffusion-like equation in isotopic space after cessation of infall. Only in the case of a linear $F$ can $f(t,x)=F(x)$ persist throughout evolution. In Fig. \ref{concave vs convex}, we show how the covariation of normalized isotopic contribution evolves when $x$ and $X$ obey the parabolic convex and parabolic concave injection functions, respectively. So the CC and EOR trends could represent a functional relationship modified from that of the parent cloud at a given epoch, two such functional relationships corresponding to two different epochs, or one or both may individually include a time evolution (similarly to that advocated by \citet{Schilleretal2018}). It would suffice that e.g. s- and r-process components had different gradient shapes in the parental cloud. So the anticorrelation between $\delta^{54}$Cr and $\delta^{50}$Ti among CCs needs not be interpreted in terms of an admixture of a high-$\delta^{50}$Ti but negative $\delta^{54}$Cr (-1 \textperthousand) refractory endmember as 
 inferred by \citet{Alexander2019CC} despite lack of evidence in the meteoritical record, as this trend may not necessarily be extrapolated to the CAI formation epoch.

\subsection{On thermal processing as a source of isotopic heterogeneities}
\label{Thermal processing}

One feature of our model is that the isotopic composition of a rock can be identified with that of the local reservoir. This, as we have seen at the end of Section \ref{CAI}, must break down for volatile elements such as O. A widespread school of thought in the cosmochemical community    
would also deny it for refractory elements since it ascribes isotopic anomalies in bulk meteorites to ``thermal processing'' (e.g. in the CAI-forming region) of originally isotopically uniform (CI/CR-like) material \citep[e.g.][]{Niemeyer1988,HussLewis1995,Trinquieretal2009,VanKootenetal2016,Olsenetal2016,Worshametal2019}. Specifically, presolar grains would have been differentially evaporated during high-temperature episodes, so that the residual solids and the vapor would have acquired isotopic signatures different from the starting material. This scenario might have been inspired by stepwise dissolution experiments of chondrite samples by progressively harsher acids, which have long been used to constrain the nature and isotopic signatures of the contained presolar grains \citep[e.g.][]{Rotaruetal1992,DauphasSchauble2016}. However, unlike the leachates in these experiments, it is not so easy to dispose of the evaporated matter in the protoplanetary disk.

  In fact, it is even doubtful that presolar grains still existed when evaporation of the elements of interest 
   started, for diffusion during ``pre-heating'' may have long destroyed their isotopic individuality in the precursor dust aggregates. In ordinary chondrites, presolar grains are no longer identifiable for parent body metamorphism beyond petrographic type 3.8 \citep{HussLewis1995,Brandonetal2005}, corresponding to a peak temperature of $\sim$ 800 K \citep[e.g.][]{Hussetal2006}. If, for a rough estimate, the relevant diffusion coefficient followed an Arrhenian dependence with an activation energy equal to the dry mineral average of 278 kJ/mol quoted by \citet{BradyCherniak2010}, the Ma timescale spent at this temperature by these chondrites \citep{Gailetal2014} would translate into a mere $10^{5\pm 1}$ s (around a day) around 1500 K to erase isotopic heterogeneities
. 
  
    If the intra-rock heterogeneities did survive until evaporation, of course this evaporation should not have been complete if it was to leave any isotopic imprint. It may be noted though that nucleosynthetic anomalies of similar (relative) magnitudes have been seen in CAIs for elements of widely different volatilities, including Zn and Ni \citep{DauphasSchauble2016}. \citet{Nanneetal2019} also deemed unlikely that thermal stability would have favored the same nucleosynthetic component (e.g. the r-process one for the CC components) for all these elements, despite the different carrier minerals relevant for each of them. Assuming partial evaporation is not enough, for isotopic exchange between the residual solid and the vapor phase would have to be prevented e.g. by rapid cooling. We would then expect superimposed (linearly) mass-dependent fractionations, in the direction of heavy isotope enrichment for the residue (most visible for isotope systems least affected by nucleosynthetic anomalies). 
   While FUN inclusions (by definition) show isotopic mass-dependent fractionations e.g. for Mg, Si or O (even if their chemistry or mineralogy do not systematically differ from regular CAIs; \citet{Krotetal2014}), PLACs which display the largest anomalies do not \citep{Ireland1988}, indicating significant isotopic equilibration with the gas. They also show no correlation between nucleosynthetic anomalies and rare earth element patterns which may trace their the condensation ``prehistory'' \citep{Faheyetal1987,Irelandetal1988}. Conversely, no ultrarefractory CAI compiled by \citet{Ivanovaetal2012} shows any Ti isotopic fractionation.

  
  Since the isotopic effect of thermal processing for refractory elements would mostly reside on refractory residues, the isotopic variations of bulk chondrites would be expected to reflect varying admixtures of CAIs. Yet, Ni isotopic measurements by \citet{Nanneetal2019} ruled out explaining e.g. CC isotopic signatures by simple CAI accumulation in an EOR-like host \citep{Nanneetal2019}---the hosts themselves had to be CC-like to a substantial fraction, and thus hosts across the chondrite spectrum must have also varied in isotopic composition. There is, it is true, a correlation between isotopic signatures of CCs and their refractory element enrichment \citep{Trinquieretal2009}, but it needs not have been causal. Indeed, the latter, related somehow to condensation and evaporation in the inner disk \citep[e.g.][]{LarimerWasson1988,Palmeetal1988,Ciesla2008
  }, certainly correlated with heliocentric distance and/or time of accretion, as the isotopic anomalies would have in our model. 
  Finally, it is worth noting that even the putative ``unprocessed'' chondrites (CI, CR) of \citet{VanKootenetal2016,Olsenetal2016} display some isotopic variations. Not only are the $^{54}$Cr anomalies of CR chondrites resolved from those of CIs, but CI-like clasts in brecciated meteorites 
   are themselves diverse in that respect \citep{Goodrichetal2019, Patzeketal2019}, hence a relative spread at least of $\sim$0.2 \textperthousand 
  . So a significant part of the isotopic heterogeneity cannot be correlated to thermal processing.

  
\section{Conclusion}
\label{Conclusion}

 \begin{figure}[!htbp]
 \begin{center}
 {\includegraphics[width=\columnwidth]{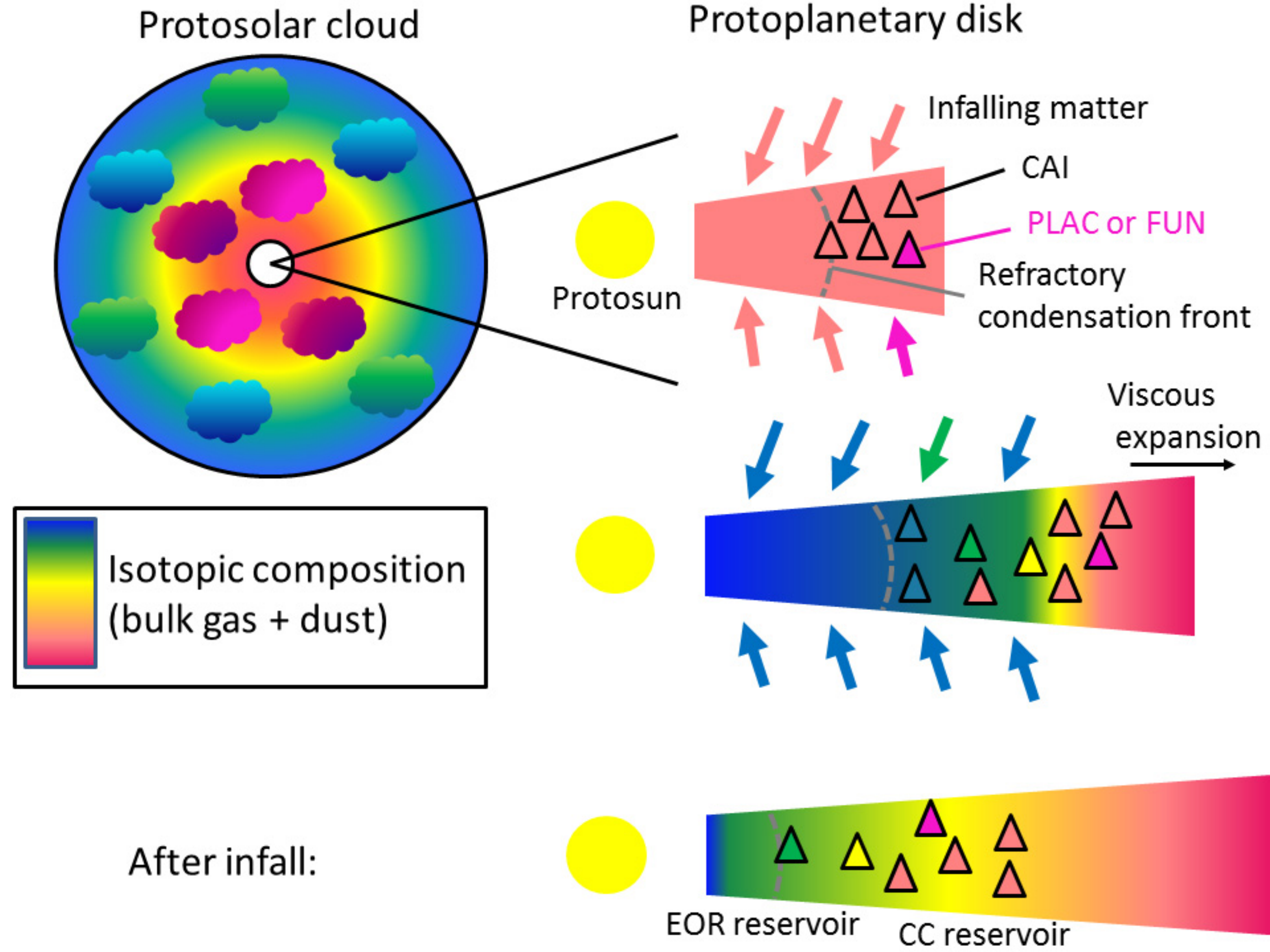}}
 \end{center}
\caption{Sketch of the scenario investigated in this paper. The color coding tracks the bulk isotopic composition of the cloud and disk but triangles single out individual CAIs. The protosolar cloud is isotopically zoned both in the radial and azimuthal directions (the latter symbolized by distinctly colored clumps, not included in our actual 1D calculations) but the disk itself is rapidly azimuthally homogenized (with only some FUN CAIs and PLACs recording the transient small-scale heterogeneity). The inner parts of the cloud collapse first and pass on their isotopic signature (r-process-enriched) to the forming CAIs which are then partly entrained outward during the viscous expansion of the disk. This reverses the isotopic gradient in the disk. CAIs continue to be produced afterward, with later isotopic signatures, but at a reduced rate, so they tend to record somewhat more ``archaic'' isotopic signatures (longer-wavelength in the color coding) than their surroundings which approximate the composition of the chondrites in which they will be incorporated.
 } 
\label{sketch}
\end{figure}

We have simulated the building and evolution of the solar protoplanetary disk as a result of the sequential collapse of a (monotonically) isotopically zoned dense core. The isotopic signatures (originally carried by presolar grains of diverse origins) were assumed to be passively transported, with no nuclear reaction or gas-solid fractionation. We find that the post-infall disk, as sampled by bulk meteorites, retains a solid-weighted standard deviation about one order of magnitude below that of the original cloud. Refractory inclusions should show systematically earlier signatures, hereby probing the infall stage of Solar System formation, so it can be inferred that the r-process component, seen to be enriched in them, tended to decrease over time in the infalling matter. While most refractory inclusions show an isotopic range about as expected from the current bulk meteorite dispersion, some such as platy hibonite crystals scatter too widely to be accounted for by our 1D model. These we speculate fossilize transient, smaller-scale heterogeneities in the protosolar cloud, superimposed on the secular evolution of the infalling matter.

  The preferential preservation of earlier bulk isotopic signatures in the outer disk, owing to its outward expansion, is consistent with the presence of carbonaceous chondrite parent bodies (which are isotopically closest to refractory inclusions) in the outer part of the main belt and beyond. How their \textit{chemical} characteristics with respect to their non-carbonaceous counterparts arose remains to be investigated, and will be the subject of a future work. The isotopic hiatus between the two superclans is also not reproduced by the model. Although Jupiter may have played a role, the exact relevant mechanism and timing remain unclear.
  
    Our model of inheritance of isotopic heterogeneities from the protosolar cloud is at variance with the popular scenario of thermal processing of an isotopically uniform primeval dust. We have however argued that presolar grains would have long disappeared before evaporation, and that CAIs show little correlation of nucleosynthetic signatures and evaporation history and cannot anyway explain alone the isotopic diversity of 
 \textit{bulk} meteorites. Still, our modelling assumptions probably break down for nonrefractory elements such as oxygen whose isotopic anomalies may have arisen in the disk.  Nevertheless, the findings of this study are likely applicable to a large class of isotopic systems---those which retain our presolar nucleosynthetic memory.

\bibliographystyle{aa}
\bibliography{bibliography}

\acknowledgments
The authors wish to acknowledge the financial support of ANR-15-CE31-0004-1 (ANR CRADLE). Prof. Nicolas Dauphas is thanked for providing isotopic data for our figures. We thank the reviewer for his/her various suggestions of improvement of the accessibility of this paper.

\newpage
\appendix

\section{Results for injection functions of Paper I}
\label{Paper I injection}

In this appendix, we display the evolution of average bulk and condensate composition for the two injection functions (monotonic with a plateau and spike) studied in Paper I. Note that these simulations were run assuming a constant fragmentation velocity of 10 m/s, unlike the runs of this paper, hence the modest lag with which the disk composition follows the injection function.

 \begin{figure}[!htbp]
 \begin{center}
 {\includegraphics[width=0.5\columnwidth]{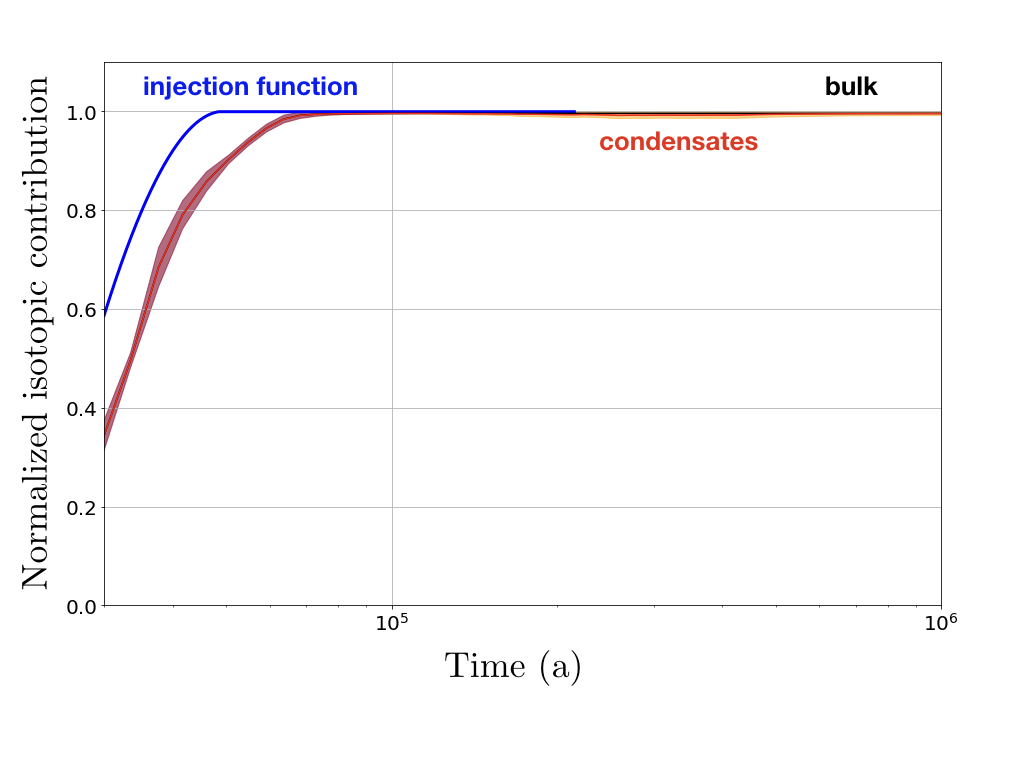}}
 {\includegraphics[width=0.5\columnwidth]{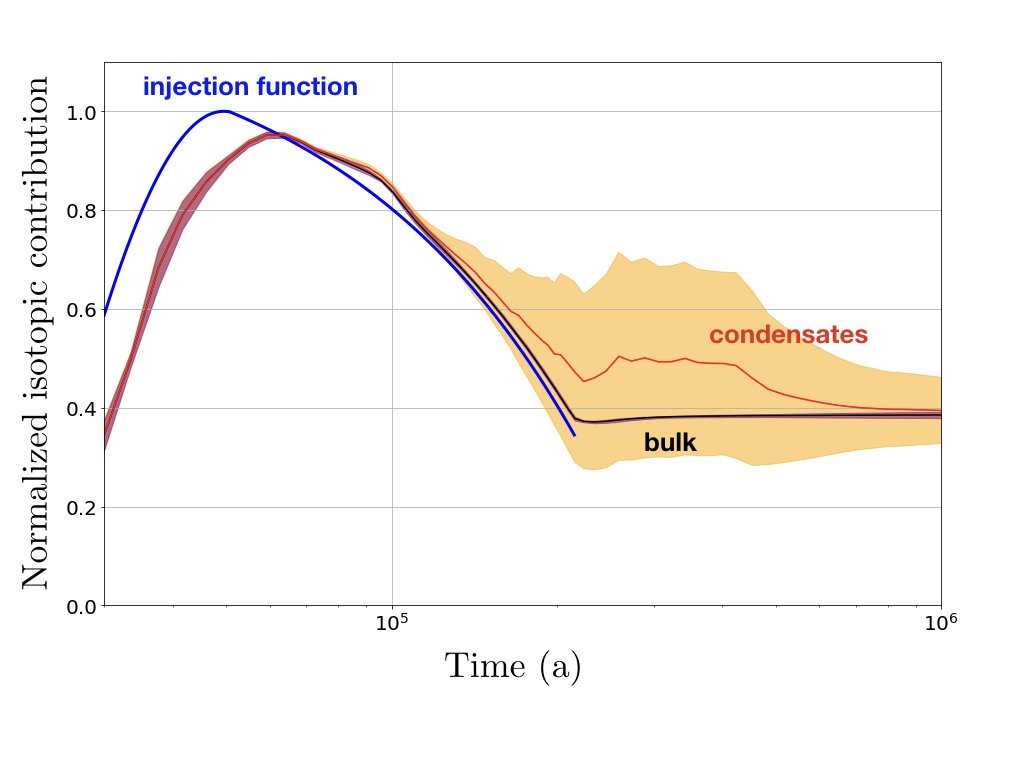}}
 \end{center}
\caption{Same as Fig. \ref{avg stdev} but for the injection functions of Paper I (overplotted) and assuming a constant fragmentation velocity of 10 m.s$^{-1}$.
 } 
\label{Paper I avg stdev}
\end{figure}

%

\vspace{5mm}

\newpage




\end{document}